\documentclass[english,twocolumn,prX,aps,showpacs,amsmath,amssymb,showkeys,superscriptaddress,nofootinbib]{revtex4}
\pdfoutput=1 
\usepackage[T1]{fontenc}
\usepackage[latin9]{inputenc}
\usepackage{color}
\usepackage{babel}

\usepackage{graphicx}
\usepackage[unicode=true, 
 bookmarks=true,bookmarksnumbered=false,bookmarksopen=false,
 breaklinks=false,pdfborder={0 0 1},backref=false,colorlinks=true]
 {hyperref}
\hypersetup{pdftitle={Scalable arrays of doped silicon RF Paul traps},
 pdfauthor={J. Britton, D. Leibfried, J. Beall, R. B. Blakestad, J. H. Wesenberg and D. J. Wineland},
 pdfsubject={ion trapping}}
 
\makeatletter
\@ifundefined{textcolor}{}
{%
 \definecolor{BLACK}{gray}{0}
 \definecolor{WHITE}{gray}{1}
 \definecolor{RED}{rgb}{1,0,0}
 \definecolor{GREEN}{rgb}{0,1,0}
 \definecolor{BLUE}{rgb}{0,0,1}
 \definecolor{CYAN}{cmyk}{1,0,0,0}
 \definecolor{MAGENTA}{cmyk}{0,1,0,0}
 \definecolor{YELLOW}{cmyk}{0,0,1,0}
 }

\usepackage{natbib}
\usepackage{bm}
\bibpunct{}{}{,}{s}{}{}

\makeatother

\begin{document}

\title{Scalable arrays of RF Paul traps in degenerate Si}

\author{J.~Britton}

\email{britton@nist.gov}

\affiliation{Quantum Electrical Metrology Division, NIST, Boulder, CO 80305, USA}

\author{D.~Leibfried}

\affiliation{Time and Frequency Division, NIST, Boulder, CO 80305, USA}

\author{J.~Beall}

\affiliation{Quantum Electrical Metrology Division, NIST, Boulder, CO 80305, USA}

\author{R.~B.~Blakestad}

\affiliation{Time and Frequency Division, NIST, Boulder, CO 80305, USA}

\author{J.~H.~Wesenberg}

\affiliation{Department of Materials, University of Oxford, Oxford OX1 3PH, UK}

\author{D.~J.~Wineland}

\affiliation{Time and Frequency Division, NIST, Boulder, CO 80305, USA}
\begin{abstract}
We report techniques for the fabrication of multi-zone linear RF Paul
traps that exploit the machinability and electrical conductivity of
degenerate silicon. The approach was tested by trapping and laser
cooling $^{24}\mbox{Mg}^{+}$ ions in two trap geometries: a single-zone
two-layer trap and a multi-zone surface-electrode trap. From the measured
ion motional heating rate we determine an electric field spectral
density at the ion's position of approximately $1\times10^{-10}\,(\mbox{V}/\mbox{m})^{2}\cdot\mbox{Hz}^{-1}$
at $\omega_{z}/2\pi=1.125\,\mbox{MHz}$ when the ion lies $40\,\mu m$
above the trap surface. One application of these devices is controlled
manipulation of atomic ion qubits, the basis of one form of quantum
information processing. 
\end{abstract}



\maketitle
\preprint{APL/L09-07581}

\pagebreak{}

Much recent ion trap research is motivated by the goal of building
a useful quantum information processor (QIP) using trapped atomic
ions as qubits (two-level quantum systems). In one approach, chains
of laser-cooled ions are confined in linear radio frequency (RF) Paul
traps with segmented trap electrodes.\cite{kielpinski2002a,amini2008a,blatt2008a,hucul2008ionTransport}
Although the theoretical and experimental groundwork has been laid
for a large-scale processor, so far only simple traps utilizing up
to eight ions have been demonstrated.\cite{haffner2005a,amini2008a,blatt2008a}
Useful qubit computations could be performed using segmented electrodes
which define an array of interconnected traps capable of holding and
manipulating a large number of ions.\cite{kielpinski2002a}

Building such devices poses several microfabrication challenges. For
example, the RF potentials (typically $100$ to $500\,\mbox{V}$,
$10$ to $100$ MHz) needed for trapping suggest use of a low-loss
substrate such as sapphire to prevent heating. However, such materials
are incompatible with many through-wafer via technologies needed to
distribute potentials to hundreds of control electrodes.\cite{kim2005a,amini2008a}
Also, in vacuum, exposed dielectrics (e.g., the substrate) can accumulate
charge giving rise to unwanted stray fields at the ions' location.
High aspect-ratio electrodes can mitigate this problem but are difficult
to fabricate for our target electrode width of $\sim10\,\mu\mbox{m}$.
It is desirable that the ratio of electrode height to inter-electrode
distance ratio be greater than $2$.\cite{amini2008a} 

\begin{figure}[!t]
\centering\includegraphics[width=1\columnwidth]{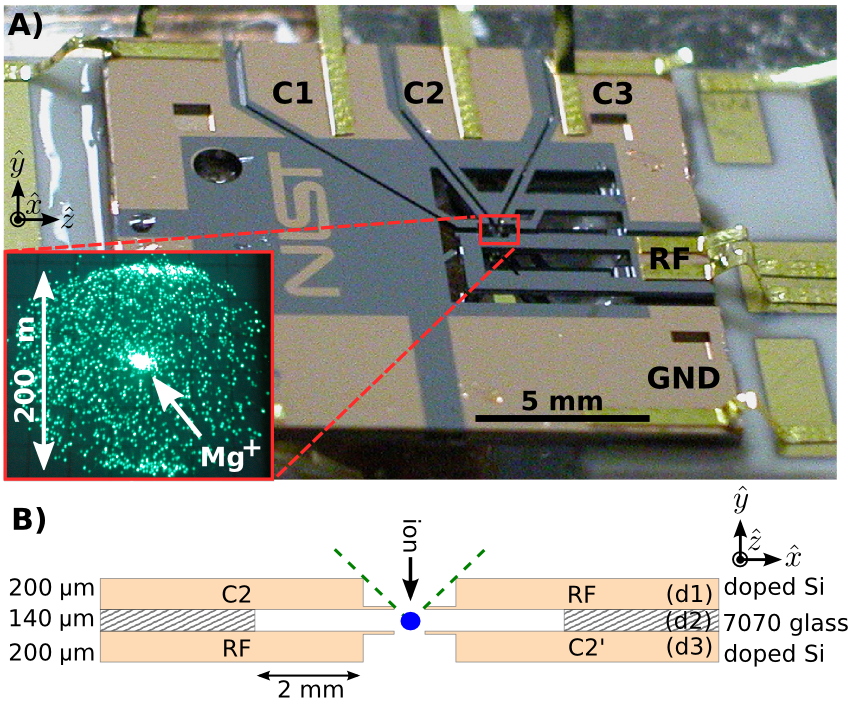} 

\caption{(Color online) A) Photograph of a two-layer silicon ion trap with
a single trapping zone. The structure consists of two layers of silicon
anodically bonded via a glass spacer. Inset: Fluorescence from a single
laser-cooled $^{24}\mbox{Mg}^{+}$ ion imaged onto a camera (viewed
from above). B) The chip geometry in cross-section near the trapping
region (indicated by a dot); not to scale. The inter-electrode spacing
was $\sim6\,\mu\mbox{m}$ (e.g., between $C_{1}$ and $C_{2}$ near
the ion). The ion-electrode distance was $\sim122\,\mu\mbox{m}$ (e.g.,
the closest distance between $C_{2}$ and the ion). The primed electrodes
are not visible in the photograph. The two-tiered etch of the silicon
trap electrodes permits a large solid angle for efficient collection
of ion fluorescence (dashed lines). }
\label{fig:2layer} 
\end{figure}

\begin{figure}[!t]
\centering\includegraphics[width=1\columnwidth]{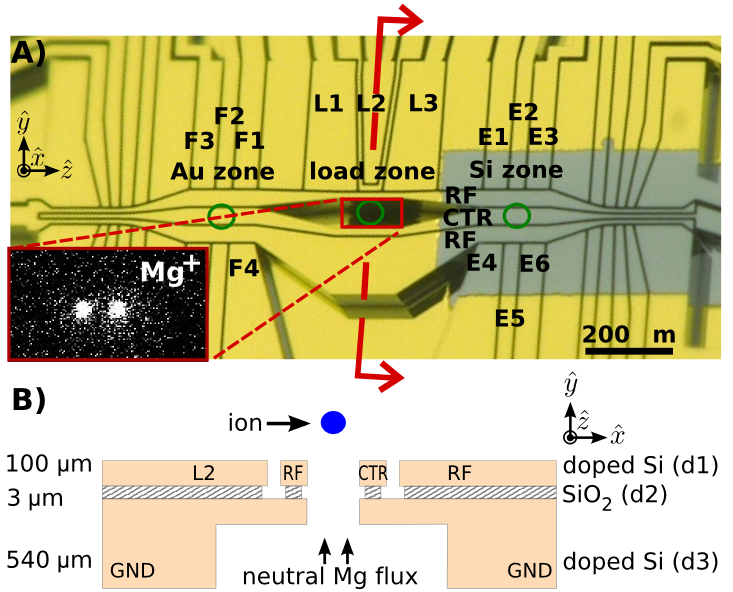}

\caption{(Color online) A) Photograph of a multi-zone surface electrode ion
trap fabricated from a single SOI wafer. Annotations highlight three
trap zones (indicated by circles): a load zone (electrodes $L_{1}-L_{3}$)
and a pair of zones whose electrode surfaces were either bare silicon
(electrodes $E_{1}-E_{6}$) or $1\,\mu\mbox{m}$ evaporated gold $F_{1}-F_{6}$,
not all labels visible). Inset: fluorescence from a pair of laser-cooled
$^{24}\mbox{Mg}^{+}$ ion imaged onto a CCD camera (viewed from above).
B) Chip geometry in cross-section near the loading zone; not to scale.
The inter-electrode spacing was $\sim4\,\mu\mbox{m}$ (e.g., between
$E_{1}$ and $E_{2}$). The SiO$_{2}$ was undercut about $\sim2\,\mu\mbox{m}$
by a wet etch. A hole cut in layer $d_{3}$ permits passage of neutral
$^{24}\mbox{Mg}$ from the back side of the wafer to the trap load
zone without risk of shorting trap electrodes. The structural insulator
($\mbox{SiO}_{2}$) was outside the field of view of the ions in all
zones. }
\label{fig:1layer} 
\end{figure}

Using degenerate silicon%
\footnote{A semiconductor is said to be \textit{degenerate} when the number
of electrons in its conduction band approaches that of a metal.%
} as an electrode material \cite{kielpinski2001a} and deep reactive
ion etching of thru-wafer channels, we have constructed structures
that meet these objectives (Figs.~\ref{fig:2layer} and \ref{fig:1layer}).
In one device (Fig.~\ref{fig:1layer}) ions were trapped above the
surface of a photolithographically patterned silicon-on-insulator
(SOI) heterostructure. Compared with manually assembled traps, this
approach reduces alignment errors.\cite{rowe2002a,blakestad2008a} 

The devices were characterized by use of trapped and laser-cooled
atomic $^{24}\mbox{Mg}^{+}$ ions. We measured the rate at which ions
gain kinetic energy from noisy ambient electric fields, an important
characteristic for QIP. \cite{turchette2000a,blatt2008a} Also, we
obtain reasonable agreement between simulation of the trap potentials
and experimentally measured frequencies, important for reliable transport
of ions between zones.\cite{amini2008a,blakestad2008a} 

The suitability of degenerate silicon as a trap electrode material
was demonstrated by building and testing several ion traps. Two of
these traps are discussed in this paper.

Figure~\ref{fig:2layer} shows a single-zone two-layer anodically
bonded trap that used a $7070$ borosilicate glass as a dielectric
spacer ($0.06$ loss tangent at $1\,\mbox{MHz}$).\cite{kielpinski2001a,britton2006b,britton2008b}
The device was built to demonstrate trapping using semiconducting
trap electrodes; prior traps at NIST with similar a geometry used
metal electrodes.\cite{rowe2002a} The silicon was doped by the manufacturer
with boron to give it a resistivity of $0.5-1\times10^{-3}\,\Omega\cdot\mbox{cm}$.
After the trap electrode pattern was etched, the silicon and glass
wafers were diced into chips and bonded. In this first demonstration
device, the chips were aligned by hand with the aid of a microscope.
The electrodes near the ions were bare silicon; the nearest glass
surface was more than $2\,\,\mbox{mm}$ away from the trap zone. 

The RF potential $\mbox{V}_{\mbox{RF}}=125\,\mbox{V}$ at $\Omega_{RF}/2\pi=67\,\mbox{MHz}$,
was applied with a $\lambda/4$ resonant transformer with a loaded
Q of $372$. Typical ion lifetime with laser cooling was several hours;
lifetime without laser cooling was up to $20\,\mbox{s}$.\cite{britton2006b}
The static trapping potentials were approximately $V(C_{1})=V(C_{1}^{'})=V(C_{3})=V(C_{3}^{'})=3\,\mbox{V}$
and $V(C_{2})=V(C_{2}^{'})=0\,\mbox{V}$. 

Figure~\ref{fig:1layer} shows a trap with a surface-electrode geometry.\cite{chiaverini05b,seidelin2006a}
The substrate was commercially available SOI with a Si resistivity
of $5\mbox{ to }20\times10^{-3}\,\Omega\cdot\mbox{cm}$.\cite{britton2008b}
The single-layer geometry is amenable to microfabrication, as all
the trap electrodes lie in a single plane. The trap has $45$ electrodes,
permitting translation of ions from a loading zone to a pair of arms
each with multiple trapping zones. Near the end of each arm the electrodes
taper so that the ion-electrode distance drops from $\sim45\,\mu\mbox{m}$
to $\sim10\,\mu\mbox{m}$; these zones were not employed for the results
reported here.

Trapping conditions were $\mbox{V}_{\mbox{RF}}=50\,\mbox{V}$ at $\Omega_{RF}/2\pi=67\,\mbox{MHz}$
and the RF resonant transformer loaded Q was $90$. Typical ion lifetime
with Doppler cooling was around an hour; without cooling the lifetime
was $10$~s. Trap potentials were determined numerically and designed
to null ion RF micromotion and properly orient the trap principle
axes.\cite{turchette2000a} Transport from the load zone to an experimental
zone was repeated hundreds of times without ion loss ($1$~Hz repetition
rate). The trap zone adjacent to electrodes $E_{2}$ and $E_{5}$
lies $\sim371\,\,\mu\mbox{m}$ away from the load zone and is $\sim41\,\,\mu\mbox{m}$
above the electrode surface. Typical static potentials were $V(E_{1})=0.71$~V,
$V(E_{2})=-0.58$~V, $V(E_{3})=0.20$~V, $V(E_{4})=0.20$~V, $V(E_{5})=-1.81$~V,
$V(E_{6})=0.71$~V and $V(E_{CTR})=0.11\,\mbox{V}$. For these potentials,
the frequency along the $\hat{z}\mbox{-axis}$ was measured to be
$\omega_{z}/2\pi=1.125$~MHz, in agreement with simulation to within
$4\,\%$. From simulation and measured radial frequencies ($\omega_{x}/2\pi=7.80$~MHz,
$\omega_{y}/2\pi=9.25$~MHz), the RF potential was inferred. The
trap depth predicted from simulation was $25$~meV. Neither device
exhibited breakdown for RF potentials up to $150\,\mbox{V}$. 

In the surface-electrode trap, kinetic energy gained by ions in the
absence of cooling was measured by use of a Doppler recooling technique.
\cite{epstein2007b,wesenberg2007a} As a test of the effect of electrode
material type on ion motional heating, two regions were created: one
with gold-coated electrodes and one with bare silicon electrodes (Fig.~\ref{fig:1layer}).
Heating measurements were made over each zone in a static potential
well at an ion-electrode distance of $\sim40\,\mu\mbox{m}$. We observed
that ion motional heating in the absence of laser cooling\cite{turchette2000a}
was the same in the gold and bare silicon experimental zones. The
inferred electric field noise spectral density was competitive with
other room temperature microtraps and was determined to be approximately
$1\times10^{-10}\,(\mbox{V}/\mbox{m})^{2}\cdot\mbox{Hz}^{-1}$ at
$\omega_{z}/2\pi=1.125\,\mbox{MHz}$ ($\sim20\times10^{3}$ quanta/s).\cite{turchette2000a,deslauriers2004a,epstein2007b,amini2008a}
Because the rate of energy gain was about the same in both zones and
somewhat variable from day to day, we suspect the heating was caused
in part by noise injected from an unidentified external source.\cite{britton2008b,blakestad2008a} 

Two types of Si-insulator-Si heterostructures were used for these
traps: anodically bonded Si-glass-Si, and commercial SOI. In both,
trap electrodes were electrically isolated islands of conducting silicon
formed by selective removal of material (Bosch DRIE). The etch pattern
was defined by $\sim7~\mu\mbox{m}$ photoresist. The narrowest practical
channel width in a $200\,\mu\mbox{m}$ wafer was $4~\mu\mbox{m}$
($50:1$ aspect-ratio). Note that typical wet etch techniques for
silicon (KOH, EDP) are ineffective in degenerate silicon.\cite{tong1999a}

Two-tiered etching (Fig.~\ref{fig:2layer}B) was accomplished by
use of a pair of overlapping etch masks and the additivity of the
etch process. The lower mask was $\sim1~\mu\mbox{m}$ thermal oxide,
the top was $\sim7\,\mu\mbox{m}$ photoresist. After an initial deep
etch, hydrogen fluoride (HF) was used to etch the exposed oxide, leaving
the photoresist intact. A second deep etch completed the two-tier
structure.\cite{britton2008b}


Si-glass-Si heterostructures were assembled by use of anodic bonding.\cite{tong1999a}
In our experiments we fabricated structures with $d_{1}=100\mbox{ to }200~\mu\mbox{m}$,
$d_{2}=140\mbox{ to }200~\mu\mbox{m}$ and $d_{3}=100\mbox{ to }600~\mu\mbox{m}$.
Anodic bonding of full wafers and individual chips was performed on
a hotplate ($450^{\circ}$~C, $500$~V). Prior to bonding, chips
were cleaned with HF and Piranha etch. 

Through-wafer features were etched in the glass spacers before anodic
bonding by ultrasonic milling. Alternately, it may be possible to
etch the glass after bonding by use of silicon as a mask for a glass
etchant such as HF. For HF:H$_{2}$O = 1:10, the vertical etch rate
is reported to be $10~\mu\mbox{m}/\mbox{hr}$ with an undercut of
$15\,\mu\mbox{m}/\mbox{hr}$.\cite{corman1998a} 

Commercial SOI wafers are available with $d_{1}=0.1\mbox{ to }500\,\mu\mbox{m}$,
$d_{2}=0.1\mbox{ to }10~\mu\mbox{m}$, $d_{3}=100\mbox{ to }500~\mu\mbox{m}$
and a resistivity as low as $0.5\times10^{-3}~\Omega\cdot\mbox{cm}$,
which helps reduce RF loss. For SOI, thermal silicon oxide usually
forms the insulating layer $(d_{2})$.

Ion trap chips were packaged in either co-fired ceramic chip carriers
or planar ceramic circuit boards, both high vacuum compatible ($\sim1\times10^{-9}\,\mbox{Pa}$)
and with gold traces.\cite{stick2007a} The chips were adhered to
the chip carriers by a ceramic paste. Ohmic contacts were deposited
on the silicon chip ($10\,\mbox{nm}$~Al, $10\,\mbox{nm}$~Ti, $1000\,\mbox{nm}$~Au).
Native oxide was stripped by a plasma etch or HF dip before deposition
of ohmic contacts and again immediately before inserting a finished
chip into the trap vacuum system.

The $^{24}\mbox{Mg}^{+}$ ions were created by electron bombardment
or resonant photo-ionization of thermally evaporated neutral magnesium
atoms. The atom source was isotopically enriched $^{24}\mbox{Mg}$
packed inside a resistively-heated stainless tube with an aperture
pointing toward the trap loading zones. The ions were cooled to the
Doppler limit by use of a laser tuned below the $^{24}\mbox{Mg}^{+}$~D2
transition at $280\,\mbox{nm}$.\cite{epstein2007b,britton2008b} 

This letter presents approaches to building ion trap arrays that use
degenerate silicon as an electrode material. Characteristics of two
trap structures were presented and their performance suggests that
this technology is compatible with trapping large arrays of ions for
applications in QIP. The planar surface-electrode device demonstrated
the feasibility of building monolithic ion traps in SOI. In addition
to multi-zone traps, the goal of quantum computing could potentially
be advanced by integration of these types of devices with CMOS electronics
(e.g., via bump bonding), MEMS optics and optical fibers.\cite{kim2005a}

We thank J.~Bollinger and J.~Amini for comments on the manuscript.
This work was supported by IARPA and the NIST Quantum Information
Program. J.~H.~W. thanks the Danish Research Agency for financial
support. This Letter is a contribution of NIST and not subject to
U.S. copyright.


\end{document}